\documentstyle[12pt,twoside, epsf, here, psfig]{article}
\newcounter{tony}
\newcommand{\bc}{\mbox{\boldmath{$c$}}}
\newcommand{\bd}{\mbox{\boldmath{$d$}}}

\newcommand{\bn}{\mbox{\boldmath{$n$}}}

\newcommand{\bt}{\mbox{\boldmath{$t$}}}

\newcommand{\bv}{\mbox{\boldmath{$v$}}}

\newcommand{\bx}{\mbox{\boldmath{$x$}}}
\newcommand{\by}{\mbox{\boldmath{$y$}}}
\newcommand{\bz}{\mbox{\boldmath{$z$}}}

\newcommand{\bI}{\mbox{\boldmath{$I$}}}

\newcommand{\bN}{\mbox{\boldmath{$N$}}}
\newcommand{\bO}{\mbox{\boldmath{$O$}}}
\newcommand{\bP}{\mbox{\boldmath{$P$}}}
\newcommand{\bQ}{\mbox{\boldmath{$Q$}}}

\newcommand{\bxi}{\mbox{\boldmath{$\xi$}}}

\newcommand{\bsigma}{\mbox{\boldmath{$\sigma$}}}

\newcommand{\beq}{\begin{equation}}
\newcommand{\eeq}[1]{\label{eq:#1}\end{equation}}

\newcommand{\eqref}[1]{(\ref{eq:#1})}

      \newcommand{\beqn}{\begin{equation}}
      \newcommand{\eeqn}{\end{equation}}
      \newcommand{\beqna}{\begin{eqnarray}}
      \newcommand{\eeqna}{\end{eqnarray}}

\title{Forces at the Sea Bed using a Finite Element Solution of the Mild Slope Wave Equation}

\renewcommand{\thefootnote}{\fnsymbol{footnote}}
\author{S. J. Childs \\ {\small\em Department of Pure and Applied
Mathematics, Rhodes University, Grahamstown,} \\ {\small\em 6140, South
Africa} \\ \\ J. W. Gonsalves \\ {\small\em Department of Mathematics and Applied Mathematics, University of Port Elizabeth,} \\ {\small\em  Port Elizabeth, 6000, South Africa}}

\renewcommand{\thefootnote}{\arabic{footnote}}

\date{}

\begin{document}

\maketitle

\begin{abstract}{\em  \noindent An algorithm to compute forces at the sea
bed from a finite element solution to the mild slope wave equation is
devised in this work. The algorithm is best considered as consisting of
two logical parts: The first is concerned with the computation of the
derivatives to a finite element solution, given the associated mesh; the
second is a bi--quadratic least squares fit which serves to model the sea
bed locally in the vicinity of a node. The force at the sea bed can be
quantified in terms of either lift and drag, the likes of Stokes' formula
or traction. While the latter quantity is the most desireable, the direct
computation of tractions at the sea bed is controversial in the context of
the mild slope wave equation as a result of the irrotationality implied by
the use of potentials. This work ultimately envisages a ``Monte Carlo''
approach using wave induced forces to elucidate presently known heavy
mineral placer deposits and, consequently, to predict the existance of
other deposits which remain as yet undiscovered.} \end{abstract}


\section{Introduction} \label{39}

The mild slope wave equation is a model for break water diffraction,
reflection and refraction which has been used with considerable success for
the quantitative prediction of ocean dynamics in a great variety of
circumstances (see {Booij} \cite{b:1} for limitations). The
model is linearised, assumes the sea bed to be locally flat, uses
potential theory and there is no turbulence (see {\sc Berkhoff}
\cite{b:3}, {\sc Bettess} and {Zienkiewicz} \cite{b:2}, {\sc Gonsalves}
\cite{g:1}). Despite this, a remarkable resemblance between the geometries
of some heavy mineral placer deposits and those of computer--generated
wave height envelopes (predicted using the mild slope wave equation for
waves moving over fairly simple, idealised bathymetries) is documented in
{\sc Childs} and {\sc Shillington} \cite{me:5}. Wave reflection,
refraction, diffraction and resonance would appear to have played a major
concentrating role in the formation of these deposits. 

An algorithm to compute forces at the sea bed from a finite element
solution to the mild slope wave equation and the associated mesh is
devised in this work. Two main components are fundamental to the logic of
the algorithm. One is concerned with the computation of the derivatives to
a finite element solution, given the associated mesh; the other is a
bi--quadratic least squares fit which serves to model the sea bed locally
in the vicinity of a node. There is a considerable advantage in developing
a routine to compute the derivatives separate from the existing code
(adapting the code to an alternative wave model would be one example). The
computation of the wave number using a Newton--Raphson scheme and other
components essential to the algorithm are also discussed.

This work ultimately envisages a ``Monte Carlo'' approach using wave
induced forces to elucidate presently known heavy mineral placer deposits
and, consequently, to predict the existance of other deposits which remain
as yet undiscovered. The intention is therefore to use the results in an empirical or qualitative (as opposed to quantitative) manner.

\subsection{Traction and the Boundary Layer Controversy}

The flow forces at the sea bed can be quantified in terms of either lift
and drag, Stokes' formula or traction. While the latter is most desireable
in physical terms, the direct computation of traction at the sea bed is
controversial in the context of the mild slope wave equation as a result
of the irrotationality implied by the use of potentials and the consequent
lack of a thorough treatment of the boundary layer. Computing the traction
indirectly (by using the solution to the mild slope wave equation
as a boundary condition in a model more suited to boundary layer
application eg. {\sc Childs} \cite{me:1}, \cite{me:3}, \cite{me:2} and
\cite{me:4}), though not impossible, is computationally exhorbitant. The
aforementioned controversy, practicality and the observed negligeable
effect of the pressure gradient on the mechanical character of fluid
motion in the vicinity of the bed (Yalin \cite{yalin:1}) suggest that
velocity\footnotemark[1] \footnotetext[1]{to which lift and drag are
squarely proportional} might be the more attractive option. Stokes'
formula is probably the most conventional option advocated by classical
texts such as {\sc Landau} and {\sc Lifshitz} \cite{l:1}. A comparative
study involving all four approaches is ultimately what is required. 

The traction formulae are by far the most complicated and they incorporate
all the elements necessary for the calculation of the other quantities
mentioned. Lift, drag and the quantities necessary to evaluate Stokes'
formula are all incidental to the traction calculation and it is for this
reason that the traction algorithm is supplied as the central theme to this work. 

This work is also concerned with the stability of fairly small, sediment
grains, grains whose threshold is presently reached at deep to
intermediate wave depths where the orbitals are relatively small. Scaling
arguments suggest that an oscillatory flow in which oscillations are
relatively small in comparison to the wave length is a potential flow to
first approximation. The lateral extent of the sediment deposits of
interest, taken in conjunction with observations that the convective term
is negligeable (Yalin \cite{yalin:1}), suggests a fairly uniform boundary
layer may be assumed. It may therefore be possible to ignore the exact
physics of the boundary layer at the scale on which the sediments of
interest occur, leaving the way open for the qualitative use of a traction
calculated directly from the solution of the mild slope wave equation.
Under these circumstances the tractional flow driving, what is assumed to
be a relatively thin and uniform boundary layer is what is being
considered. The modelled motion for a linear sea bed would be that of a
number of layers of fluid slapping up and down, a kind of pumping action
on the sea bed.

\section{Stress in Terms of a Solution to the Mild Slope Wave Equation}

The approximated velocity potential based on the solution to the mild
slope wave equation is
\begin{eqnarray} \label{137}
\Phi(x_1,x_2,x_3,t) = \mathop{\rm Re} \left\{ f^h(x_1,x_2) e^{-i
\omega t} \right\} Z(x_3,h)
\end{eqnarray}
where $\Phi$ is the velocity potential, ${\rm Re}\{ \ \}$ indicates the
real part of a complex number, $f^h$ is the finite element solution to
the mild slope wave equation, $Z$ is a function which describes
attenuation with depth, $x_3$ is the vertical coordinate measured from
mean water level, $h$ is the depth below mean water level and $\omega$
is a frequency. The stress tensor is given by the constitutive relation
\[
{\bsigma} = - p{\bI}  + \mu( \nabla {\bv}  
+ (\nabla {\bv} )^t ),
\]
where, in terms of the approximation (\ref{137}),
\renewcommand{\thefootnote}{\fnsymbol{footnote}}
\begin{eqnarray*}
v_{1,1} &=& \mathop{\rm Re} \left\{ \left( \frac{\partial^2
f^h}{\partial x^2_1} Z + 2 \frac{\partial f^h}{\partial x_1}
\frac{\partial Z}{\partial x_1} + f^h \frac{\partial^2 Z}{\partial
x^2_1} \right) e^{-i \omega t} \right\} \\
v_{2,2} &=& \mathop{\rm Re} \left\{ \left( \frac{\partial^2
f^h}{\partial x^2_2} Z + 2 \frac{\partial f^h}{\partial x_2}
\frac{\partial Z}{\partial x_2} + f^h \frac{\partial^2 Z}{\partial
x^2_2} \right) e^{-i \omega t} \right\} \\ 
v_{3,3} &=& \mathop{\rm Re} \left\{ \left( f^h \frac{\partial^2
Z}{\partial x^2_3} \right) e^{-i \omega t} \right\} \\
v_{1,2} &=& \footnotemark[2] v_{2,1} \ = \ \mathop{\rm Re} \left\{
\left( \frac{\partial^2 f^h}{\partial x_2 \partial x_1} Z +
\frac{\partial f^h}{\partial x_2} \frac{\partial Z}{\partial x_1} +
\frac{\partial f^h}{\partial x_1} \frac{\partial Z}{\partial x_2} + f^h
\frac{\partial^2 Z}{\partial x_2 \partial x_1} \right) e^{-i \omega t}
\right\} \\ 
v_{1,3} &=& \footnotemark[2] v_{3,1} \ = \ \mathop{\rm Re} \left\{
\left(\frac{\partial f^h}{\partial x_1} \frac{\partial Z}{\partial x_3}
+ f^h \frac{\partial^2 Z}{\partial x_3 \partial x_1} \right) e^{-i
\omega t} \right\} \\
v_{2,3} &=& \footnotemark[2] v_{3,2} \ = \ \mathop{\rm Re} \left\{
\left(\frac{\partial f^h}{\partial x_2} \frac{\partial Z}{\partial x_3}
+ f^h \frac{\partial^2 Z}{\partial x_3 \partial x_2} \right) e^{-i
\omega t} \right\}.
\end{eqnarray*}
Forthcoming sections are devoted to the modelling and computation of
these values. \footnotetext[2]{Notice that the symmetry of the stress
tensor is preserved when introducing the approximation.}
\renewcommand{\thefootnote}{\arabic{footnote}}

\section{The Analytic Derivatives of a Finite Element Solution}
\label{16}

The finite element method approximates a solution to a problem in a
finite dimensional subspace $\bar F^h$. Thus for $f^h \in \bar F^h$, 
\[
f^h({\bx} ) = \sum^{\scriptsize \mbox{\it nPoint}}_{i=1} c_i \psi_i({\bx} ) 
\]
where $\mbox{\it nPoint}$ is the total number of nodes, the $c_i$'s are
the degrees of freedom (the discrete solution) and the $\psi_i({\bx}
)$'s are the shape functions. The local approximation on each element
is
\[
f^h({\bx}) \mid_{\Omega_e} = \sum^{\scriptsize \mbox{\it nNode}}_{i=1}
c_i^{(e)} \psi_i^{(e)}({\bx}),
\]
where $\mbox{\it nNode}$ is the number of nodes per element, the
$c_i^{(e)}$'s are the local degrees of freedom, the $\psi_i^{(e)}({\bx}
)$'s are the localised shape functions and $\Omega_e$ is the element in
question. Differentiating both sides of the above equation,
\begin{eqnarray} \label{10}
\left. \frac{\partial^j f^h}{\partial x_k \cdots \partial x_l}
\right|_{\Omega_e} &=& \sum^{\scriptsize \mbox{\it nNode}}_{i=1}
c_i^{(e)} \frac{\partial^j \psi_i^{(e)}}{\partial x_k \cdots \partial
x_l}.
\end{eqnarray}
The problem of calculating the derivatives of a finite element solution
therefore translates directly into one of calculating the derivatives of
the localised shape functions on each element. These localised shape functions are defined in terms of a basis as follows
\[
\psi_i^{(e)}({{\bx} ({\bxi})}) \equiv \phi_i({\bxi}).
\]
where the $\{\phi_i(\bxi)\}$ is the basis defined on the master element domain, $\hat \Omega$. In this way the problem can be transferred into one in terms of the master element.

\subsection{The Two--Dimensional Case}

For a two dimensional problem
\begin{eqnarray} \label{12}
\left[ \frac{\partial f^h}{\partial x_1}, \ \frac{\partial
f^h}{\partial x_2}, \ \frac{\partial^2 f^h}{\partial x_1^2},
\ \frac{\partial^2 f^h}{\partial x_2^2}, \ \frac{\partial^2
f^h}{\partial x_1 \partial x_2} \right]
&=& \sum^{\scriptsize \mbox{\it nNode}}_{i=1} c_i^{(e)} \left[
\frac{\partial \psi_i^{(e)}}{\partial x_1}, \ \frac{\partial
\psi_i^{(e)}}{\partial x_2}, \ \frac{\partial^2 \psi_i^{(e)}}{\partial
x_1^2}, \ \frac{\partial^2 \psi_i^{(e)}}{\partial x_2^2},
\ \frac{\partial^2 \psi_i^{(e)}}{\partial x_1 \partial x_2} \right]
\nonumber \\
& & \hspace{30mm} \mbox{\it (by equation (\ref{10})).}
\end{eqnarray} 
Applying the chain rule the first derivative of the basis with respect to
the first variable is
\begin{eqnarray*}
\frac{\partial \phi_i}{\partial \xi_1} &=& \frac{\partial
\psi_i^{(e)}}{\partial x_1} \frac{\partial x_1}{\partial \xi_1} +
\frac{\partial \psi_i^{(e)}}{\partial x_2} \frac{\partial x_2}{\partial
\xi_1} \\ &=& \left[ \displaystyle \frac{\partial x_1}{\partial \xi_1},
\ \displaystyle \frac{\partial x_2}{\partial \xi_1} \right] \left[
\begin{array}{c} \displaystyle \frac{\partial \psi_i^{(e)}}{\partial
x_1} \\ \\ \displaystyle \frac{\partial \psi_i^{(e)}}{\partial x_2}
\end{array} \right].
\end{eqnarray*} 
The first derivative of the basis with respect to the second variable
is
\begin{eqnarray*}
\frac{\partial \phi_i}{\partial \xi_2} &=& \frac{\partial
\psi_i^{(e)}}{\partial x_1} \frac{\partial x_1}{\partial \xi_2} +
\frac{\partial \psi_i^{(e)}}{\partial x_2} \frac{\partial x_2}{\partial
\xi_2} \\ &=& \left[ \displaystyle \frac{\partial x_1}{\partial \xi_2},
\ \displaystyle \frac{\partial x_2}{\partial \xi_2} \right] \left[
\begin{array}{c} \displaystyle \frac{\partial \psi_i^{(e)}}{\partial
x_1} \\ \\ \displaystyle \frac{\partial \psi_i^{(e)}}{\partial x_2}
\end{array} \right].
\end{eqnarray*} 
The second derivative of the basis with respect to the
first variable is 
\begin{eqnarray*}
\frac{\partial^2 \phi_i}{\partial \xi_1^2}
&=& \frac{\partial}{\partial \xi_1} \left\{ \frac{\partial
\psi_i^{(e)}}{\partial x_1} \frac{\partial x_1}{\partial \xi_1} +
\frac{\partial \psi_i^{(e)}}{\partial x_2} \frac{\partial x_2}{\partial
\xi_1} \right\} \\
&=& \frac{\partial^2 \psi_i^{(e)}}{\partial x_1^2} \left(
\frac{\partial x_1}{\partial \xi_1} \right)^{2}  + \frac{\partial^2
\psi_i^{(e)}}{\partial x_2 \partial x_1} \frac{\partial x_1}{\partial
\xi_1} \frac{\partial x_2}{\partial \xi_1} + \frac{\partial
\psi_i^{(e)}}{\partial x_1} \frac{\partial^2  x_1}{\partial \xi_1^2 } +
\frac{\partial^2 \psi_i^{(e)}}{\partial x_2^2} \left( \frac{\partial
x_2}{\partial \xi_1}  \right)^{2} \\
& & + \frac{\partial^2 \psi_i^{(e)}}{\partial x_1 \partial x_2}
\frac{\partial x_1}{\partial \xi_1} \frac{\partial x_2}{\partial \xi_1}
+ \frac{\partial \psi_i^{(e)}}{\partial x_2} \frac{\partial^2  x_2}{\partial
\xi_1^2 } \\
&=& \left[
\displaystyle \frac{\partial^2 x_1}{\partial \xi_1^2}, \ \displaystyle
\frac{\partial^2 x_2}{\partial \xi_1^2}, \ \left( \displaystyle
\frac{\partial x_1}{\partial \xi_1} \right)^2, \ \left( \displaystyle
\frac{\partial x_2}{\partial \xi_1} \right)^2, \ 2 \displaystyle
\frac{\partial x_1}{\partial \xi_1} \frac{\partial x_2}{\partial \xi_1}
\right]
\left[ \begin{array}{c} 
\displaystyle \frac{\partial \psi_i^{(e)}}{\partial x_1} \\ \\ 
\displaystyle \frac{\partial \psi_i^{(e)}}{\partial x_2} \\ \\ 
\displaystyle \frac{\partial^2 \psi_i^{(e)}}{\partial x_1^2} \\ \\ 
\displaystyle \frac{\partial^2 \psi_i^{(e)}}{\partial x_2^2} \\ \\ 
\displaystyle \frac{\partial^2 \psi_i^{(e)}}{\partial x_1 \partial x_2}
\end{array} \right].
\end{eqnarray*} 
The second derivative of the basis with respect to the
second variable is 
\begin{eqnarray*}
\frac{\partial^2 \phi_i}{\partial \xi_2^2}
&=& \frac{\partial}{\partial \xi_2} \left\{ \frac{\partial
\psi_i^{(e)}}{\partial x_1} \frac{\partial x_1}{\partial \xi_2} +
\frac{\partial \psi_i^{(e)}}{\partial x_2} \frac{\partial x_2}{\partial
\xi_2} \right\} \\
&=& \frac{\partial^2 \psi_i^{(e)}}{\partial x_1^2} \left(
\frac{\partial x_1}{\partial \xi_2} \right)^{2}  + \frac{\partial^2
\psi_i^{(e)}}{\partial x_2 \partial x_1} \frac{\partial x_1}{\partial
\xi_1} \frac{\partial x_2}{\partial \xi_2} + \frac{\partial
\psi_i^{(e)}}{\partial x_1} \frac{\partial^2  x_1}{\partial \xi_2^2} +
\frac{\partial^2 \psi_i^{(e)}}{\partial x_2^2} \left( \frac{\partial
x_2}{\partial \xi_2}  \right)^{2} \\
& & + \frac{\partial^2 \psi_i^{(e)}}{\partial x_1 \partial x_2}
\frac{\partial x_1}{\partial \xi_2} \frac{\partial x_2}{\partial \xi_2}
+ \frac{\partial \psi_i^{(e)}}{\partial x_2} \frac{\partial^2
x_2}{\partial \xi_2^2 } \\
&=& \left[
\displaystyle \frac{\partial^2 x_1}{\partial \xi_2^2}, \ \displaystyle
\frac{\partial^2 x_2}{\partial \xi_2^2}, \ \left( \displaystyle
\frac{\partial x_1}{\partial \xi_2} \right)^2, \ \left( \displaystyle
\frac{\partial x_2}{\partial \xi_2} \right)^2, \ 2 \displaystyle
\frac{\partial x_1}{\partial \xi_2} \frac{\partial x_2}{\partial \xi_2}
\right]
\left[ \begin{array}{c} 
\displaystyle \frac{\partial \psi_i^{(e)}}{\partial x_1} \\ \\ 
\displaystyle \frac{\partial \psi_i^{(e)}}{\partial x_2} \\ \\ 
\displaystyle \frac{\partial^2 \psi_i^{(e)}}{\partial x_1^2} \\ \\ 
\displaystyle \frac{\partial^2 \psi_i^{(e)}}{\partial x_2^2} \\ \\ 
\displaystyle \frac{\partial^2 \psi_i^{(e)}}{\partial x_1 \partial x_2} 
\end{array} \right].
\end{eqnarray*} 
The cross derivative of the basis is 
\begin{eqnarray*}
\frac{\partial^2 \phi_i}{\partial \xi_1 \partial \xi_2} &=&
\frac{\partial}{\partial \xi_1} \left\{ \frac{\partial
\psi_i^{(e)}}{\partial x_1} \frac{\partial x_1}{\partial \xi_2} +
\frac{\partial \psi_i^{(e)}}{\partial x_2} \frac{\partial x_2}{\partial
\xi_2} \right\} \\ &=& \frac{\partial^2 \psi_i^{(e)}}{\partial x_1^2}
\frac{\partial x_1}{\partial \xi_1} \frac{\partial x_1}{\partial \xi_2}
+ \frac{\partial^2 \psi_i^{(e)}}{\partial x_2 \partial x_1}
\frac{\partial x_1}{\partial \xi_2} \frac{\partial x_2}{\partial \xi_1}
+ \frac{\partial \psi_i^{(e)}}{\partial x_1} \frac{\partial^2
x_1}{\partial \xi_1 \partial \xi_2} + \frac{\partial^2
\psi_i^{(e)}}{\partial x_2^2} \frac{\partial x_2}{\partial \xi_1}
\frac{\partial x_2}{\partial \xi_2} \\ & & + \frac{\partial^2
\psi_i^{(e)}}{\partial x_1 \partial x_2} \frac{\partial x_1}{\partial
\xi_1} \frac{\partial x_2}{\partial \xi_2} + \frac{\partial
\psi_i^{(e)}}{\partial x_2} \frac{\partial^2  x_2}{\partial \xi_1
\partial \xi_2} \\
&=& \left[
\displaystyle \frac{\partial^2 x_1}{\partial \xi_1 \partial \xi_2},
\ \displaystyle \frac{\partial^2 x_2}{\partial \xi_1 \partial \xi_2},
\ \displaystyle \frac{\partial x_1}{\partial \xi_1} \frac{\partial
x_1}{\partial \xi_2}, \ \displaystyle  \frac{\partial x_2}{\partial
\xi_1} \frac{\partial x_2}{\partial \xi_2}, \ \left( \displaystyle
\frac{\partial x_1}{\partial \xi_1} \frac{\partial x_2}{\partial \xi_2}
+ \frac{\partial x_1}{\partial \xi_2} \frac{\partial x_2}{\partial
\xi_1} \right)
\right]
\left[ \begin{array}{c} 
\displaystyle \frac{\partial \psi_i^{(e)}}{\partial x_1} \\ \\ 
\displaystyle \frac{\partial \psi_i^{(e)}}{\partial x_2} \\ \\ 
\displaystyle \frac{\partial^2 \psi_i^{(e)}}{\partial x_1^2} \\ \\ 
\displaystyle \frac{\partial^2 \psi_i^{(e)}}{\partial x_2^2} \\ \\ 
\displaystyle \frac{\partial^2 \psi_i^{(e)}}{\partial x_1 \partial x_2}
\end{array} \right].
\end{eqnarray*} 
Collecting the above expressions together and re--expressing them in
a vector--matrix form,
\begin{eqnarray} \label{14}
\underbrace{\left[ \begin{array}{c} 
\displaystyle \frac{\partial \phi_i}{\partial \xi_1} \\ \\ 
\displaystyle \frac{\partial \phi_i}{\partial \xi_2} \\ \\ 
\displaystyle \frac{\partial^2 \phi_i}{\partial \xi_1^2} \\ \\ 
\displaystyle \frac{\partial^2 \phi_i}{\partial \xi_2^2} \\ \\ 
\displaystyle \frac{\partial^2 \phi_i}{\partial \xi_1 \partial \xi_2} 
\end{array} \right]}_{{\bd}^i({\bxi})}
&=& \underbrace{\left[ \begin{array}{ccccc} 
\displaystyle \frac{\partial x_1}{\partial \xi_1}  &  \displaystyle
\frac{\partial x_2}{\partial \xi_1} & 0 & 0 & 0 \\ \\ \displaystyle
\frac{\partial x_1}{\partial \xi_2}  &  \displaystyle \frac{\partial
x_2}{\partial \xi_2} & 0 & 0 & 0 \\ \\ \displaystyle \frac{\partial^2
x_1}{\partial \xi_1^2}  &  \displaystyle \frac{\partial^2 x_2}{\partial
\xi_1^2} & \left( \displaystyle \frac{\partial x_1}{\partial \xi_1}
\right)^2  & \left( \displaystyle \frac{\partial x_2}{\partial \xi_1}
\right)^2 & 2 \displaystyle \frac{\partial x_1}{\partial \xi_1}
\frac{\partial x_2}{\partial \xi_1} \\ \\ \displaystyle
\frac{\partial^2 x_1}{\partial \xi_2^2}  &  \displaystyle
\frac{\partial^2 x_2}{\partial \xi_2^2} & \left( \displaystyle
\frac{\partial x_1}{\partial \xi_2} \right)^2  & \left( \displaystyle
\frac{\partial x_2}{\partial \xi_2} \right)^2 & 2 \displaystyle
\frac{\partial x_1}{\partial \xi_2} \frac{\partial x_2}{\partial \xi_2}
\\ \\ \displaystyle \frac{\partial^2 x_1}{\partial \xi_1 \partial
\xi_2}  &  \displaystyle \frac{\partial^2 x_2}{\partial \xi_1 \partial
\xi_2} & \displaystyle \frac{\partial x_1}{\partial \xi_1}
\frac{\partial x_1}{\partial \xi_2} & \displaystyle  \frac{\partial
x_2}{\partial \xi_1} \frac{\partial x_2}{\partial \xi_2} & \left(
\displaystyle \frac{\partial x_1}{\partial \xi_1} \frac{\partial
x_2}{\partial \xi_2} + \frac{\partial x_1}{\partial \xi_2}
\frac{\partial x_2}{\partial \xi_1} \right)
\end{array} \right]}_{{\bQ}({\bxi})}
\left[ \begin{array}{c} 
\displaystyle \frac{\partial \psi_i^{(e)}}{\partial x_1} \\ \\ 
\displaystyle \frac{\partial \psi_i^{(e)}}{\partial x_2} \\ \\ 
\displaystyle \frac{\partial^2 \psi_i^{(e)}}{\partial x_1^2} \\ \\ 
\displaystyle \frac{\partial^2 \psi_i^{(e)}}{\partial x_2^2} \\ \\ 
\displaystyle \frac{\partial^2 \psi_i^{(e)}}{\partial x_1 \partial x_2} 
\end{array} \right]. \nonumber \\
& & 
\end{eqnarray} 
It follows from equation (\ref{14}) that
\[
\left. \left[ \frac{\partial \psi_i^{(e)}}{\partial x_1}, \
\frac{\partial \psi_i^{(e)}}{\partial x_2}, \ \frac{\partial^2
\psi_i^{(e)}}{\partial x_1^2}, \ \frac{\partial^2
\psi_i^{(e)}}{\partial x_2^2}, \ \frac{\partial^2
\psi_i^{(e)}}{\partial x_1 \partial x_2} \right] \right|_{{\bx}
({\bxi})} = \left[ {\bQ}({\bxi}) \right]^{-1} {\bd}^i({\bxi}).
\]
This equation is the formula by which the much desired shape function
derivatives are calculated. Substituting it into equation (\ref{12})
\begin{eqnarray} \label{13} 
\left. \left[ \frac{\partial f^h}{\partial x_1}, \ \frac{\partial
f^h}{\partial x_2}, \ \frac{\partial^2 f^h}{\partial x_1^2},
\ \frac{\partial^2 f^h}{\partial x_2^2}, \ \frac{\partial^2
f^h}{\partial x_1 \partial x_2} \right] \right|_{{\bx}({\bxi})} &=&
\sum^{\scriptsize \mbox{\it nNode}}_{i=1} c_i^{(e)} \left[
{\bQ}({\bxi}) \right]^{-1} {\bd} ^i({\bxi}).
\end{eqnarray}
This equation is the formula by which the first, second and cross
derivatives of a finite element solution to a two--dimensional problem
are calculated.

{\bf The Matrix Entries:} The matrix entries may all be formulated by
taking derivatives of the finite element mapping. Taking the
opportunity to develop a systematic notation for the purposes of the
algorithm simultaneously,
\begin{eqnarray} \label{139}
x_i(\bxi) = \sum_{k=1}^{\scriptsize \mbox{\it nNode}} \phi_k(\bxi)
(x_i\mid_{node \ k}) &=& \sum_{k=1}^{\scriptsize \mbox{\it nNode}}
\mbox{\it shape}(k,1) * \mbox{\it eCoord}(k,i)
\end{eqnarray}
where $(x_i\mid_{node \ k})$ is the $i$th coordinate of node
$k$, as is {\it eCoord}$(k,i)$, the $\phi_k(\bxi)$'s
are the basis, as are the {\it shape}$(k,1)$'s. The matrix entries are
calculated according to
\begin{eqnarray} \label{140}
\frac{\partial x_i}{\partial \xi_j} &=& \ \sum_{k=1}^{\scriptsize
\mbox{\it nNode}} \frac{\partial \phi_k}{\partial \xi_j} (x_i\mid_{node
\ k}) \ = \ \sum_{k=1}^{\scriptsize \mbox{\it nNode}} \mbox{\it
shape}(k,j+1) * \mbox{\it eCoord}(k,i) \nonumber \\
\frac{\partial^2 x_i}{\partial \xi_j^2} &=& \sum_{k=1}^{\scriptsize
\mbox{\it nNode}} \frac{\partial^2 \phi_k}{\partial \xi_j^2}
(x_i\mid_{node \ k}) \ = \ \sum_{k=1}^{\scriptsize \mbox{\it nNode}}
\mbox{\it shape}(k,j+3) * \mbox{\it eCoord}(k,i) \nonumber \\
\frac{\partial^2 x_i}{\partial \xi_1 \partial \xi_2} &=&
\sum_{k=1}^{\scriptsize \mbox{\it nNode}} \frac{\partial^2
\phi_k}{\partial \xi_1 \partial \xi_2} (x_i\mid_{node \ k}) \ =
\ \sum_{k=1}^{\scriptsize \mbox{\it nNode}} \mbox{\it shape}(k,6) *
\mbox{\it eCoord}(k,i),
\end{eqnarray}
where the definition of the {\it shape}$(k,j)$'s follows from
the equations above.

{\bf The Derivatives of the Basis:} Obtaining formulae for the various
derivatives of the basis is an elementary exercise in differentiation. The
resulting formulae in the particular instance of the 8--noded
quadrilateral basis (appendix, page \pageref{9}) are listed in the
appendix on page \pageref{17}. A combined structure--flow chart diagram of
the algorithm which computes the derivatives of a finite element solution
is given on page \pageref{88}.

\begin{figure}
\begin{center} \leavevmode
\mbox{\epsfbox{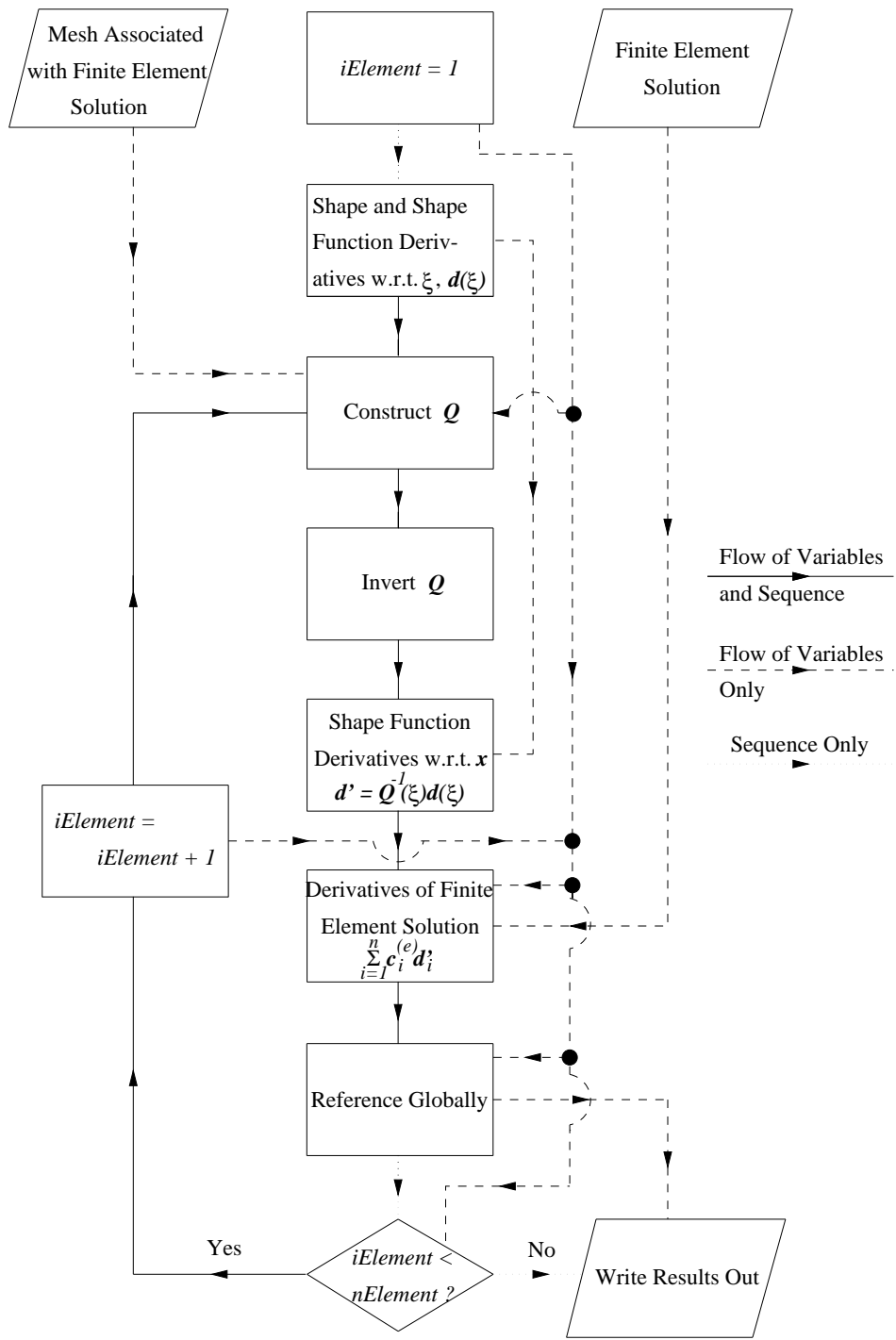}}
\end{center}
\caption{Combined Structure--Flow Chart Diagram of an Algorithm which
Computes the Derivatives of a Finite Element Solution Analytically.}
\label{88}
\end{figure}

\subsection{Some Test Examples}

\begin{figure}[H]
\begin{center} \leavevmode
\mbox{\epsfbox{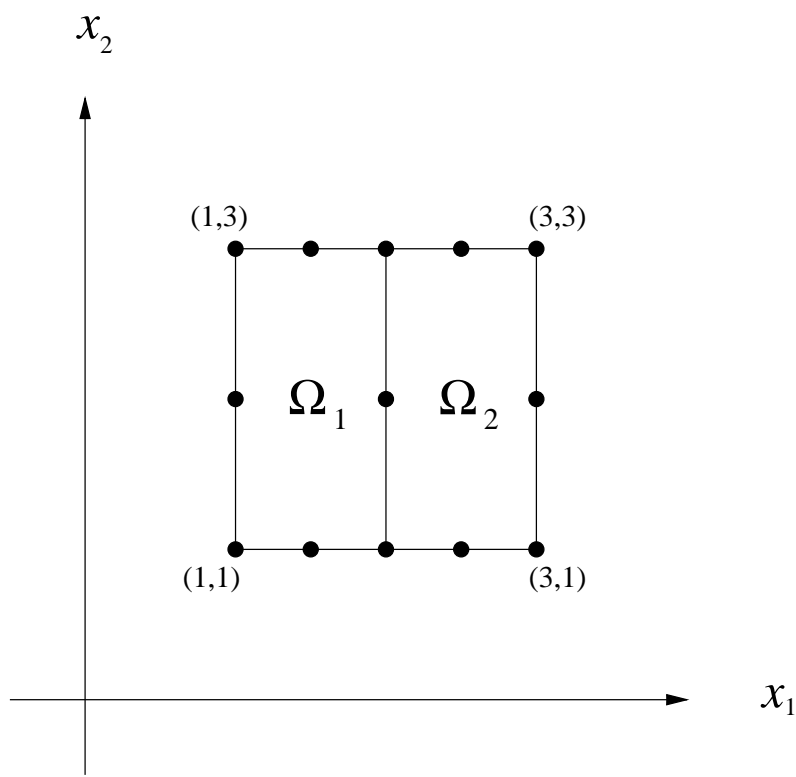}}
\end{center}
\caption{Test Mesh} \label{168}
\end{figure}

Finite element approximations for a number of simple, analytic surfaces
were devised by evaluating the self same functions at the nodes of the
test mesh depicted in Figure \ref{168}. A comparison of the various
derivatives of the approximated surface with those of the analytic
function itself confirmed the algorithm to be working.

{\bf Test 1:} For the surface 
\[
f(x_1, x_2) = 1,
\]
the first, second and crossed derivatives were obtained to specified
precision (approximately 16 significant figures) at all thirteen
nodes.

{\bf Test 2:} For the surface 
\[
f(x_1, x_2) = x_1,
\]
the first, second and crossed derivatives were obtained to specified
precision (approximately 16 significant figures) at all thirteen
nodes.

{\bf Test 3:} For the surface 
\[
f(x_1, x_2) = x_1^2 - 4x_1 + 3
\]
the first, second and crossed derivatives were obtained to specified
precision (approximately 16 significant figures) at all thirteen
nodes.

\section{The Various Derivatives of $Z(x_3,h)$}

The function which describes the attenuation with
depth is
\[
Z(x_3,h) = \frac{\cosh(\kappa(h + x_3))}{\cosh(\kappa h)}
\]
where the $x_3$ coordinate is measured from the mean water level, $h$
is the depth below mean water level, $\pi$ is the usual mathematical
constant and $\kappa$ is defined by the non--dimensional dispersion
relation
\[
\frac{1}{\kappa} = {\mathop{\rm Tanh}}(\kappa h).
\]
Observing that $h = h(x_1, x_2)$ and $\kappa = \kappa(x_1, x_2)$, the
first, second and cross derivatives are accordingly formulated in the
appendix on page \pageref{18}. At the sea bed where $x_3 = - h$:
\begin{eqnarray*}
Z \mid_{x_3 = -h} 
&=& \frac{1}{\cosh(\kappa h)}, \\
& & \\
\left. \frac{\partial Z}{\partial x_1} \right |_{x_3 = -h} &=&
\frac{-1}{\cosh(\kappa h)} \left( \frac{\partial
h}{\partial x_1} + \frac{h}{\kappa} \frac{\partial \kappa}{\partial
x_1} \right), \\
& & \\
\left. \frac{\partial Z}{\partial x_2} \right |_{x_3 = -h} &=&
\frac{-1}{\cosh(\kappa h)} \left( \frac{\partial
h}{\partial x_2} + \frac{h}{\kappa} \frac{\partial \kappa}{\partial
x_2} \right), \\
& & \\
\left. \frac{\partial Z}{\partial x_3} \right |_{x_3 = -h} &=& 0, \\
& & \\
\left. \frac{\partial^2 Z}{\partial x_1^2} \right |_{x_3 = -h} &=&
\frac{1}{\cosh(\kappa h)} \left[ (1 + \kappa^2) \left(
\frac{\partial h}{\partial x_1} \right)^2 - \frac{\partial^2
h}{\partial x_1^2}  + h \left( 2 \kappa - 1 - {\sinh(\kappa
h)} - \frac{1}{\kappa} \right. \right. \\ & & \hspace{25mm} \left.
\left. - \frac{1}{h \kappa} \right) \frac{\partial h}{\partial x_1}
\frac{\partial \kappa}{\partial x_1} + \frac{h}{\kappa} \left(
\frac{2}{\kappa} + h \kappa \right) \left( \frac{\partial
\kappa}{\partial x_1} \right)^2 - \frac{h}{\kappa} \frac{\partial^2
\kappa}{\partial x_1^2} \right], \\
& & \\
\left. \frac{\partial^2 Z}{\partial x_2^2} \right |_{x_3 = -h} &=&
\frac{1}{\cosh(\kappa h)} \left[ (1 + \kappa^2) \left(
\frac{\partial h}{\partial x_2} \right)^2 - \frac{\partial^2
h}{\partial x_2^2}  + h \left( 2 \kappa - 1 - {\sinh(\kappa
h)} - \frac{1}{\kappa} \right. \right. \\ & & \hspace{25mm} \left.
\left. - \frac{1}{h \kappa} \right) \frac{\partial h}{\partial x_2}
\frac{\partial \kappa}{\partial x_2} + \frac{h}{\kappa} \left(
\frac{2}{\kappa} + h \kappa \right) \left( \frac{\partial
\kappa}{\partial x_2} \right)^2 - \frac{h}{\kappa} \frac{\partial^2
\kappa}{\partial x_2^2} \right], \\
& & \\
\left. \frac{\partial^2 Z}{\partial x_1 \partial x_2} \right |_{x_3 =
-h} &=& \frac{1}{\cosh(\kappa h)} \left[ \left( 1 +
\kappa^2 \right) \frac{\partial h}{\partial x_1}  \frac{\partial
h}{\partial x_2} - \frac{\partial^2 h}{\partial x_1 \partial x_2} +
\left( h \kappa - \frac{h}{\kappa} - \frac{1}{\kappa} \right) \right.
\\
& & \\
& & \left. \left( \frac{\partial h}{\partial x_1} \frac{\partial
\kappa}{\partial x_2} + \frac{\partial h}{\partial x_2} \frac{\partial
\kappa}{\partial x_1} \right) + \left( \frac{2 h}{\kappa^2} + h^2
\right) \frac{\partial \kappa}{\partial x_1} \frac{\partial
\kappa}{\partial x_2} - \frac{h}{\kappa} \frac{\partial^2
\kappa}{\partial x_1 \partial x_2} \right], \\
& & \\
\left. \frac{\partial^2 Z}{\partial x_1 \partial x_3} \right |_{x_3 =
-h} &=& \frac{\kappa}{\cosh(\kappa h)} \left( \kappa
\frac{\partial h}{\partial x_1} + h \frac{\partial \kappa}{\partial
x_1} \right), \\
& & \\
\left. \frac{\partial^2 Z}{\partial x_2 \partial x_3} \right |_{x_3 =
-h} &=& \frac{\kappa}{\cosh(\kappa h)} \left( \kappa
\frac{\partial h}{\partial x_2} + h \frac{\partial \kappa}{\partial
x_2} \right) \hspace{10mm} \mbox{and} \\
& & \\
\left. \frac{\partial^2 Z}{\partial x_3^2} \right |_{x_3 = -h} 
&=& \frac{\left( \kappa \right)^2}{\cosh(\kappa h)}
\end{eqnarray*}

\section{The Nodal Values of $\kappa(x_1,x_2)$ and its Various
Derivatives} 

Calculating the wave number, $\kappa$, for a given depth is standard
procedure. The dispersion relation
\[
\frac{1}{\kappa} = {\mathop{\rm Tanh}}(\kappa h)
\]
is conventionally solved using Newton's method. The resulting iterative
scheme is,
\[
\kappa^{i + 1} = \kappa^{i} - \frac{\kappa^i \mathop{\rm Tanh}(\kappa^i
h) - 1}{\mathop{\rm Tanh}(\kappa^i h) + h \kappa^{i} (1 - \mathop{\rm
Tanh}^2(\kappa^i h))}
\]
where the superscript $i$ denotes the successive iteration from which a
given solution was obtained. The initial guess usually taken is
\[
\kappa = \frac{2 \pi}{\lambda_0},
\]
where $\lambda_0$ is deep water wave--length.  

Once this has been accomplished for each of the $n$ nodes belonging to
a given element, there is no reason why these nodal values shouldn't be
regarded as a discrete solution in order to determine the derivatives.
Substituting into equation (\ref{13})
\begin{eqnarray*} 
\left. \left[ \frac{\partial \kappa^h}{\partial x_1}, \ \frac{\partial
\kappa^h}{\partial x_2}, \ \frac{\partial^2 \kappa^h}{\partial x_1^2},
\ \frac{\partial^2 \kappa^h}{\partial x_2^2}, \ \frac{\partial^2
\kappa^h}{\partial x_1 \partial x_2} \right] \right|_{{\bx}({\bxi})}
&=& \sum^{\scriptsize \mbox{\it nNode}}_{j=1} \kappa \mid_{node \ j}
\left[ {\bQ}({\bxi}) \right]^{-1} {\bd}^j({\bxi}).
\end{eqnarray*}
The derivatives of $\kappa$ can, alternatively, be calculated by the
implicit differentiation of the dispersion relation. Considering
$\left[ {\bQ}({\bxi}) \right]^{-1} {\bd}^j({\bxi})$ must be calculated
at each node $j$, the former method is the more efficient.

\section{The Sea Bed at a Node}

Because nodes do not necessarily coincide with individual points of
bathymmetry measurement, and for the purposes of taking derivatives, a
``sea bed'' needs to be interpolated locally. A straightforward fit of
an $n$ degree polynomial to the $n$ data points nearest a node, the use
of cubic splines and a local least squares fit were all considered as
possible ways to interpolate bathymmetry between individual points of
bathymetry measurement.

The manner in which available data was collected proved to be a deciding
factor in the final choice. While the use of cubic splines is fairly
established in the modelling of known surfaces, the problem with
unknown surfaces is that slope information at the ``knots'' is
required.  Such information is never available in the raw bathymetry
data. A further factor to consider is that the actual data sampling
intervals range anywhere from slightly, to highly, irregular. One
advantage of the least squares method is that a large data set can be
taken into account, even individual data points weighted according to
their proximity.

The argument against fitting an $n$ degree polynomial exactly to the
nearest $n$ points in the vicinity of a given node is that the use of a
high degree polynomial will result in a totally fictitious model in cases
where the actual surface is of ``lower degree'' than the polynomial used,
alternatively, where the sampling intervals are poor.  Fitting a low
degree polynomial surface could result in the use of an unrepresentative
data sample. The solution is therefore to fit a fairly simple, low degree
polynomial surface to a larger data set. This can be accomplished using
the least squares method. A method based on the least absolute value of
the errors is preferable in theory, of course, but not in practice.

Bi--quadratic and bi--cubic surfaces were experimented with using the
method of least squares. The former was decided to be the better
choice. Irregular data was found to allow extreme cases of the
``wiggle'' effect in the bi--cubic case. A bi--cubic surface also
requires a far greater, hence locally less relevant data set and its
greater degree is therefore not necessarily an advantage. In a real--life
data comparison between actual measured depths, the depths
predicted using cubic splines and those predicted using a local, least
squares, bi--quadratic fit, a limited inspection suggested the least squares bi--quadratic fit to be superior.

\subsection{The Least Squares Fit of a Bi--Quadratic Function}

A generalised bi--quadratic equation has the form
\[
h(x,y) = c_1 + c_2y + c_3x + c_4xy + c_5y^2 + c_6x^2
\]
or when written as the dot product of two vectors,
\begin{eqnarray}\label{8} 
h(x,y) = [1, \ y, \ x, \ xy, \ y^2, \ x^2] \cdot [c_1, \ c_2, \ c_3,
\ c_4, \ c_5, \ c_6].  
\end{eqnarray}
A least squares fit makes, what is in one sense, an optimal choice of
the constants, $c_1, c_2, \cdots, c_6$. ``In one sense'', in that it
minimises the summed squares of the errors at the data points and not
the summed absolute values of these errors. The sum of the squares of
the errors, $\epsilon$, is 
\[
\epsilon = \sum_{i=1}^n (c_1 + c_2y_i + c_3x_i + c_4x_iy_i + c_5y_i^2
+ c_6x_i^2 - z_i)^2
\]
where the $z_i$ are the $n$ data points located at $(x_i, y_i)$, the
points to which the bi--quadratic equation is to be fitted. In order to
minimise $\epsilon$ with respect to the unknown constants,
\begin{eqnarray*}
\frac{\partial \epsilon}{\partial c_1} = 0 &\Rightarrow& \sum_{i=1}^n
(c_1 + c_2y_i + c_3x_i + c_4x_iy_i + c_5y_i^2 + c_6x_i^2) =
\sum_{i=1}^n z_i \\ \frac{\partial \epsilon}{\partial c_2} = 0
&\Rightarrow& \sum_{i=1}^n y_i(c_1 + c_2y_i + c_3x_i + c_4x_iy_i +
c_5y_i^2 + c_6x_i^2) = \sum_{i=1}^n y_iz_i \\ \frac{\partial
\epsilon}{\partial c_3} = 0 &\Rightarrow& \sum_{i=1}^n x_i(c_1 +
c_2y_i + c_3x_i + c_4x_iy_i + c_5y_i^2 + c_6x_i^2) = \sum_{i=1}^n
x_iz_i \\ \frac{\partial \epsilon}{\partial c_4} = 0 &\Rightarrow&
\sum_{i=1}^n x_iy_i(c_1 + c_2y_i + c_3x_i + c_4x_iy_i + c_5y_i^2 +
c_6x_i^2) = \sum_{i=1}^n x_iy_iz_i \\ \frac{\partial
\epsilon}{\partial c_5} = 0 &\Rightarrow& \sum_{i=1}^n y_i^2(c_1 +
c_2y_i + c_3x_i + c_4x_iy_i + c_5y_i^2 + c_6x_i^2) = \sum_{i=1}^n
y_i^2z_i \\ \frac{\partial \epsilon}{\partial c_6} = 0 &\Rightarrow&
\sum_{i=1}^n x_i^2(c_1 + c_2y_i + c_3x_i + c_4x_iy_i + c_5y_i^2 +
c_6x_i^2) = \sum_{i=1}^n x_i^2z_i
\end{eqnarray*}
Re--expressing the above system of equations in vector--matrix form,
\begin{eqnarray*}
\frac{\partial \epsilon}{\partial {\bc}} &=& 0
\ \Rightarrow \nonumber \\
& & \nonumber \\
\underbrace{ \sum_{i=1}^n \left[ \begin{array}{cccccc}
1 & y_i & x_i & x_iy_i & y_i^2 & x_i^2 \\ \\
y_i & y_i^2 & x_iy_i & x_iy_i^2 & y_i^3 & x_i^2y_i \\ \\
x_i & x_iy_i & x_i^2 & x_i^2y_i & x_iy_i^2 & x_i^3 \\ \\
x_iy_i & x_iy_i^2 & x_i^2y_i & x_i^2y_i^2 & x_iy_i^3 & x_i^3y_i \\ \\
y_i^2 & y_i^3 & x_iy_i^2 & x_iy_i^3 & y_i^4 & x_i^2y_i^2 \\ \\
x_i^2 & x_i^2y_i & x_i^3 & x_i^3y_i & x_i^2y_i^2 & x_i^4 
\end{array} \right] }_{\bP}
\left[ \begin{array}{c}
c_1 \\ \\ c_2 \\ \\ c_3 \\ \\ c_4 \\ \\ c_5 \\ \\ c_6 
\end {array} \right] &=&
\underbrace{ \left[ \begin{array}{cccc}
1 & 1 & \cdots & 1 \\ \\
y_1 & y_2 & \cdots & y_n \\ \\
x_1 & x_2 & \cdots & x_n \\ \\
x_1y_1 & x_2y_2 & \cdots & x_ny_n \\ \\
y_1^2 & y_2^2 & \cdots & y_n^2 \\ \\
x_1^2 & x_2^2 & \cdots & x_n^2
\end {array} \right] }_{\bO}
\left[ \begin{array}{c}
z_1 \\ \\ z_2 \\ \\ \vdots \\ \\ z_n
\end {array} \right].
\end{eqnarray*}
Therefore
\[
{\bP}{\bc} = {\bO}{\bz},
\]
where ${\bP}$ and ${\bO}$ take their respective definitions from the
previous equation. Solving for ${\bc}$,
\begin{eqnarray} \label{141}
{\bc} &=& {\bP}^{-1}{\bO}{\bz}.
\end{eqnarray}
Substituting this result into equation (\ref{8}),
\[
h(x,y) = [1, \ y, \ x, \ xy, \ y^2, \ x^2] \cdot {\bP} ^{-1}{\bO} {\bz} 
\]
where $h(x,y)$ is the depth modelled locally by this least squares
fitted, bi--quadratic equation.

\subsection{The Various Derivatives of $h(x,y)$} \label{15}

The corresponding derivatives of the sea bed are:
\begin{eqnarray*}
\frac{\partial h}{\partial x} &=& [0, \ 0, \ 1, \ y, \ 0, \ 2x] \cdot
{\bf P}^{-1}{\bO} {\bz} \\
\frac{\partial h}{\partial y} &=& [0, \ 1, \ 0, \ x, \ 2y, \ 0] \cdot
{\bP}^{-1}{\bO} {\bz} \\
\frac{\partial^2 h}{\partial x^2} &=& [0, \ 0, \ 0, \ 0, \ 0,
\ 2] \cdot {\bP}^{-1}{\bf O}{\bz}  \\ 
\frac{\partial^2 h}{\partial y^2} &=& [0, \ 0, \ 0, \ 0, \ 2, \ 0]
\cdot {\bP}^{-1}{\bO} {\bz} \\
\frac{\partial^2 h}{\partial x \partial y} &=& [0, \ 0, \ 0, \ 1, \ 0,
\ 0] \cdot {\bP}^{-1}{\bO} {\bz} 
\end{eqnarray*}

\subsection{The Sea Bed Normal}

The components of the sea bed normal are:
\begin{eqnarray*} 
N_1(x,y) &=& - \frac{\partial h}{\partial x} \ = \ - [0, \ 0, \ 1, \ y,
\ 0, \ 2x] \cdot {\bP}^{-1}{\bO} {\bz} \\ 
N_2(x,y) &=& - \frac{\partial h}{\partial y} \ = \ - [0, \ 1, \ 0, \ x,
\ 2y, \ 0] \cdot {\bP} ^{-1}{\bO}{\bz} \\
N_3(x,y) &=& \frac{\partial h}{\partial h} \ = \ 1,
\end{eqnarray*}
and the unit normal,
\[
{\bn} (x,y) = \frac{{\bN} (x,y)}{\mid \mid {\bN} (x,y) \mid \mid_2}.
\]

A combined structure--flow chart diagram of an algorithm to model the
sea bed locally in the vicinity of a node by way of a least squares
fitted bi--quadratic can be found on page \pageref{89}.

\begin{figure}
\begin{center} \leavevmode
\mbox{\epsfbox{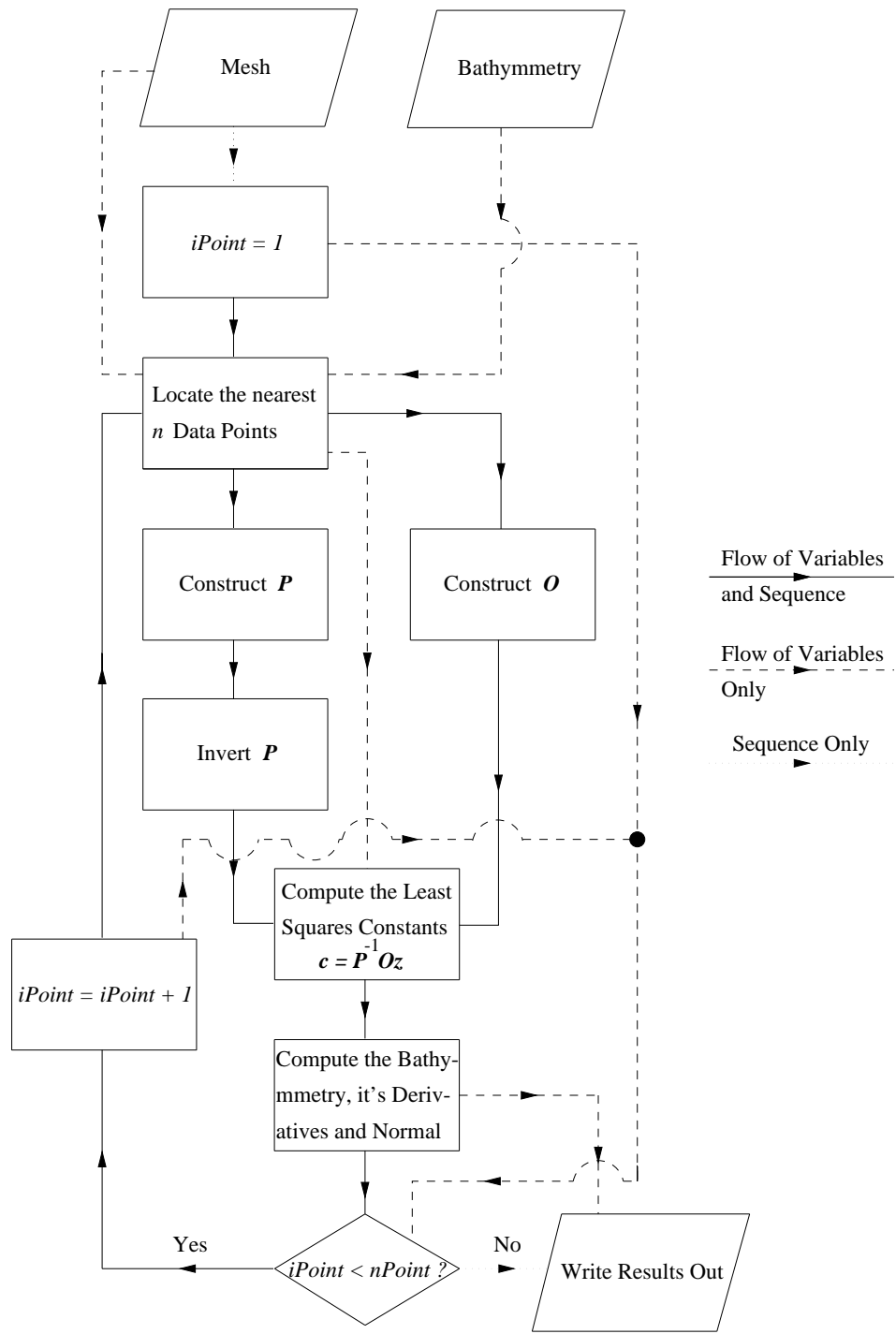}}
\end{center}
\caption{Combined Structure--Flow Chart Diagram of an Algorithm Used to
Model the Sea Bed Locally in the Vicinity of a Node.} \label{89}
\end{figure}

\subsection{Some Test Examples}

Data with which to test the algorithm was generated by evaluating a few
simple, analytic surfaces at the required number of points. A
comparison of outputted bathymetries and sea--bed normals with those of
the corresponding analytic function, from which the data was generated,
showed the algorithm to be working.

{\bf Tests 1:} The trivial cases 
\[
h(x,y) = c, \ c \ \mbox{ a constant}
\]
were used to generate the input
\begin{table}[H]
\begin{tabular}{c c c c c c c}  
$ \ \ {\bx}  \ \ $ & $ \ \ 1 \ \ $ & $ \ \ 1 \ \ $ & $ \ \ 2 \ \ $ & $
\ \ 1 \ \ $ & $ \ \ 3 \ \ $ & $ \ \ 2 \ \ $ \\
& & & & & & \\ 
{\by}  & 1 & 2 & 1 & 3 & 1 & 2 \\ 
& & & & & & \\ 
{\bz}  & c & c & c & c & c & c  
\end{tabular}.
\end{table}
The algorithm calculated both depth and normal correct to specified
precision (approximately 16 significant figures).

{\bf Test 2:} For a topography containing the arbitrarily selected
bi--quadratic
\[
h(x,y) = x^2 + 2y^2 + 3xy + 4x + 5y + 6 
\]
the input generated was \begin{table}[H]
\begin{tabular}{c c c c c c c} 
$ \ \ {\bx}  \ \ $ & $ \ \ 1 \ \ $ & $ \ \ 1 \ \ $ & $ \ \ 2 \ \ $ & $
\ \ 1 \ \ $ & $ \ \ 3 \ \ $ & $ \ \ 2 \ \ $ \\
& & & & & & \\ 
{\by}  & 1 & 2 & 1 & 3 & 1 & 2 \\ 
& & & & & & \\  
{\bz}  & 21 & 35 & 31 & 53 & 43 & 48
\end{tabular}.
\end{table}
The algorithm calculated depth and normal correct to specified
precision (approximately 16 significant figures).

{\bf Test 3:} A real--life data comparison was made between actual
measured depths, the depths predicted using a local, least squares,
bi--quadratic fit and those predicted using cubic splines. A limited
inspection suggested the least squares, bi--quadratic fit to be the
superior choice.

The algorithm is therefore considered to adequately perform the tasks
for which it was designed.

\section{The Traction Acting on the Sea Bed} \label{175}

The surface force per unit area, exerted by the fluid and acting on the
sea bed, is given by
\[
{\bt} = {\bsigma}{\bn} 
\]
where ${\bsigma}$ is the stress tensor at the sea bed and ${\bn} $ is
the unit normal to the sea bed. In terms of the quantities discussed
and formulated so far,
\begin{eqnarray*}
t_1 &=& - \ p n_1 \\
& & + \ 2 \mu \mathop{\rm Re} \left\{ e^{-i \omega t} \frac{\partial^2
f^h}{\partial x^2_1} \right\} \frac{1}{\cosh(\kappa h)} n_1
\\
& & + \ 2 \mu \mathop{\rm Re} \left\{ e^{-i \omega t} 2 \frac{\partial
f^h}{\partial x_1} \right\} \frac{-1}{\cosh(\kappa h)}
\left( \frac{\partial h}{\partial x_1} + \frac{h}{\kappa}
\frac{\partial \kappa}{\partial x_1} \right) n_1 \\
& & + \ 2 \mu \mathop{\rm Re} \left\{ e^{-i \omega t} f^h \right\}
\frac{1}{\cosh(\kappa h)} \left[ (1 + \kappa^2) \left(
\frac{\partial h}{\partial x_1} \right)^2 - \frac{\partial^2
h}{\partial x_1^2} \right. \\ 
& & \hspace{20mm} + h \left( 2 \kappa - 1 - {\sinh(\kappa
h)} - \frac{1}{\kappa} - \frac{1}{h \kappa} \right) \frac{\partial
h}{\partial x_1} \frac{\partial \kappa}{\partial x_1} \\
& & \hspace{35mm} \left. + \frac{h}{\kappa} \left( \frac{2}{\kappa} + h
\kappa \right) \left( \frac{\partial \kappa}{\partial x_1} \right)^2 -
\ \frac{h}{\kappa} \frac{\partial^2 \kappa}{\partial x_1^2} \right] n_1
\\
& & + \ 2 \mu \mathop{\rm Re} \left\{ e^{-i \omega t}
\frac{\partial^2 f^h}{\partial x_2 \partial x_1} \right\}
\frac{1}{\cosh(\kappa h)} n_2 \\
& & + \ 2 \mu \mathop{\rm Re} \left\{ e^{-i \omega t} \frac{\partial
f^h}{\partial x_2} \right\} \frac{-1}{\cosh(\kappa h)}
\left( \frac{\partial h}{\partial x_1} + \frac{h}{\kappa}
\frac{\partial \kappa}{\partial x_1} \right) n_2 \\
& & + \ 2 \mu \mathop{\rm Re} \left\{ e^{-i \omega t} \frac{\partial
f^h}{\partial x_1} \right\} \frac{-1}{\cosh(\kappa h)}
\left( \frac{\partial h}{\partial x_2} + \frac{h}{\kappa}
\frac{\partial \kappa}{\partial x_2} \right) n_2 \\
& & + \ 2 \mu \mathop{\rm Re} \left\{ e^{-i \omega t} f^h \right\}
\frac{1}{\cosh(\kappa h)} \left[ \left( 1 + \kappa^2
\right) \frac{\partial h}{\partial x_1}  \frac{\partial h}{\partial
x_2} \right. \\
& &  \hspace{20mm} - \frac{\partial^2 h}{\partial x_1 \partial x_2} +
\left( h \kappa - \frac{h}{\kappa} - \frac{1}{\kappa} \right) \left(
\frac{\partial h}{\partial x_1} \frac{\partial \kappa}{\partial x_2} +
\frac{\partial h}{\partial x_2} \frac{\partial \kappa}{\partial x_1}
\right) \\
& & \hspace{40mm} + \left( \frac{2 h}{\kappa^2} + h^2 \right) \frac{\partial
\kappa}{\partial x_1} \frac{\partial \kappa}{\partial x_2} \left. -
\frac{h}{\kappa} \frac{\partial^2 \kappa}{\partial x_1 \partial x_2}
\right] n_2 \\
& & + \ 2 \mu \mathop{\rm Re} \left\{ e^{-i \omega t} \frac{\partial^2
f^h}{\partial x_2 \partial x_1} \right\} \frac{1}{\cosh(\kappa h)} n_3 \\ 
& & + \ 2 \mu \mathop{\rm Re} \left\{ e^{-i \omega t} \frac{\partial
f^h}{\partial x_2} \right\} \frac{-1}{\cosh(\kappa h)}
\left( \frac{\partial h}{\partial x_1} + \frac{h}{\kappa}
\frac{\partial \kappa}{\partial x_1} \right) n_3 \\
& & + \ 2 \mu \mathop{\rm Re} \left\{ e^{-i \omega t} \frac{\partial
f^h}{\partial x_1} \right\} \frac{-1}{\cosh(\kappa h)}
\left( \frac{\partial h}{\partial x_2} + \frac{h}{\kappa}
\frac{\partial \kappa}{\partial x_2} \right) n_3 \\
& & + \ 2 \mu \mathop{\rm Re} \left\{ e^{-i \omega t} f^h \right\}
\frac{1}{\cosh(\kappa h)} \left[ \left( 1 + \kappa^2
\right) \frac{\partial h}{\partial x_1}  \frac{\partial h}{\partial
x_2} - \frac{\partial^2 h}{\partial x_1 \partial x_2} \right. \\
& & \hspace{20mm} + \left( h \kappa - \frac{h}{\kappa} -
\frac{1}{\kappa} \right) \left( \frac{\partial h}{\partial x_1}
\frac{\partial \kappa}{\partial x_2} + \frac{\partial h}{\partial x_2}
\frac{\partial \kappa}{\partial x_1} \right) \\
& & \hspace{20mm} \left. + \left( \frac{2 h}{\kappa^2} + h^2 \right)
\frac{\partial \kappa}{\partial x_1} \frac{\partial \kappa}{\partial
x_2} - \frac{h}{\kappa} \frac{\partial^2 \kappa}{\partial x_1 \partial
x_2} \right] n_3 \\
& & \\
t_2 &=& - \ p n_2 \\
& & + \ 2 \mu \mathop{\rm Re} \left\{ e^{-i \omega t} \frac{\partial^2
f^h}{\partial x_2 \partial x_1} \right\} \frac{1}{\cosh(\kappa h)} n_1 \\
& & + \ 2 \mu \mathop{\rm Re} \left\{ e^{-i \omega t} \frac{\partial
f^h}{\partial x_2} \right\} \frac{-1}{\cosh(\kappa h)}
\left( \frac{\partial h}{\partial x_1} + \frac{h}{\kappa}
\frac{\partial \kappa}{\partial x_1} \right) n_1 \\
& & + \ 2 \mu \mathop{\rm Re} \left\{ e^{-i \omega t} \frac{\partial
f^h}{\partial x_1} \right\} \frac{-1}{\cosh(\kappa h)}
\left( \frac{\partial h}{\partial x_2} + \frac{h}{\kappa}
\frac{\partial \kappa}{\partial x_2} \right) n_1 \\
& & + \ 2 \mu \mathop{\rm Re} \left\{ e^{-i \omega t} f^h \right\}
\frac{1}{\cosh(\kappa h)} \left[ \left( 1 + \kappa^2
\right) \frac{\partial h}{\partial x_1}  \frac{\partial h}{\partial
x_2} - \frac{\partial^2 h}{\partial x_1 \partial x_2} \right. \\
& & \hspace{20mm} + \left( h \kappa - \frac{h}{\kappa} -
\frac{1}{\kappa} \right) \left( \frac{\partial h}{\partial x_1}
\frac{\partial \kappa}{\partial x_2} + \frac{\partial h}{\partial x_2}
\frac{\partial \kappa}{\partial x_1} \right) \\
& & \hspace{20mm} \left. + \left( \frac{2 h}{\kappa^2} + h^2 \right)
\frac{\partial \kappa}{\partial x_1} \frac{\partial \kappa}{\partial
x_2} - \frac{h}{\kappa} \frac{\partial^2 \kappa}{\partial x_1 \partial
x_2} \right] n_1 \\
& & + \ 2 \mu \mathop{\rm Re} \left\{ e^{-i
\omega t} \frac{\partial^2 f^h}{\partial x^2_2} \right\}
\frac{1}{\cosh(\kappa h)} n_2 \\
& & + \ 2 \mu \mathop{\rm Re} \left\{ e^{-i \omega t} 2 \frac{\partial
f^h}{\partial x_2} \right\} \frac{-1}{\cosh(\kappa h)}
\left( \frac{\partial h}{\partial x_2} + \frac{h}{\kappa}
\frac{\partial \kappa}{\partial x_2} \right) n_2 \\
& & + \ 2 \mu \mathop{\rm Re} \left\{ e^{-i \omega t} f^h \right\}
\frac{1}{\cosh(\kappa h)} \left[ (1 + \kappa^2) \left(
\frac{\partial h}{\partial x_2} \right)^2 - \frac{\partial^2
h}{\partial x_2^2} \right. \\
& & \hspace{20mm} + h \left( 2 \kappa - 1 - {\sinh(\kappa
h)} - \frac{1}{\kappa} - \frac{1}{h \kappa} \right) \frac{\partial
h}{\partial x_2} \frac{\partial \kappa}{\partial x_2} \\
& & \hspace{35mm} \left. + \frac{h}{\kappa} \left( \frac{2}{\kappa} + h
\kappa \right) \left( \frac{\partial \kappa}{\partial x_2} \right)^2 -
\frac{h}{\kappa} \frac{\partial^2 \kappa}{\partial x_2^2} \right] n_2
\\
& & + \ 2 \mu \mathop{\rm Re} \left\{ e^{-i \omega t} f^h
\right\} \frac{\kappa}{\cosh(\kappa h)} \left( \kappa
\frac{\partial h}{\partial x_2} + h \frac{\partial \kappa}{\partial
x_2} \right) n_3 \\
& & \\
t_3 &=& - \ p n_3 \\
& & + \ 2 \mu \mathop{\rm Re} \left\{ e^{-i \omega t} f^h \right\}
\frac{\kappa}{\cosh(\kappa h)} \left( \kappa \frac{\partial
h}{\partial x_1} + h \frac{\partial \kappa}{\partial x_1} \right) n_1
\\
& & + \ 2 \mu \mathop{\rm Re} \left\{ e^{-i \omega t} f^h \right\}
\frac{\kappa}{\cosh(\kappa h)} \left( \kappa \frac{\partial
h}{\partial x_2} + h \frac{\partial \kappa}{\partial x_2} \right) n_2
\\
& & + \ 2 \mu \mathop{\rm Re} \left\{ e^{-i \omega t} f^h \right\}
\frac{\left( \kappa \right)^2}{\cosh(\kappa h)} n_3.
\end{eqnarray*}
where $p$ is the pressure, $\mu$ is the viscosity, $e$ and $i$ denote the
usual mathematical constants, $\omega$ is a frequency, $t$ is time,
$f^h$ is the finite element solution to the mild slope wave equation,
$x_3$ is the vertical coordinate measured from mean water level, $h$ is
the depth below mean water level (with the exception of the
superscript) and $\kappa$ is the wave number.  The derivatives
$\frac{\partial h}{\partial x_1}, \frac{\partial h}{\partial x_2}$
etc.  denote the $\frac{\partial h}{\partial x}, \frac{\partial
h}{\partial y}$ etc. derivatives formulated in Subection \ref{15} on page
\pageref{15} (the variables $x$ and $y$ were used in place of $x_1$ and
$x_2$ so as to avoid confusion with the first and
second data points, $(x_1, \ y_1, \ z_1)$ and $(x_2, \ y_2, \ z_2)$
respectively).

A structure chart of the entire algorithm to compute tractions on the
sea bed from a solution to the mild slope wave equation is given on page \pageref{171}.
\begin{figure}[H]
\begin{center} \leavevmode
\mbox{\epsfbox{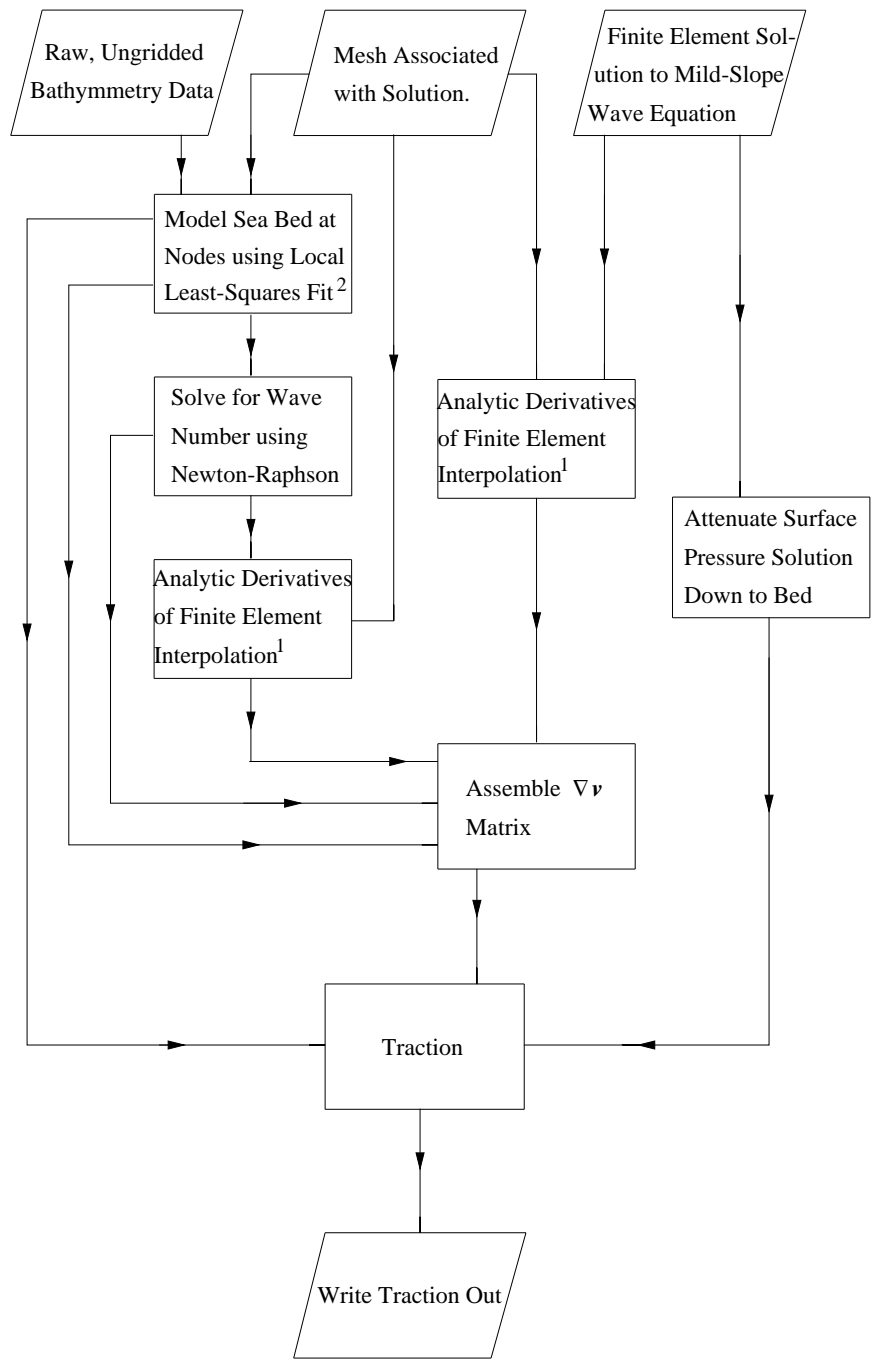}}
\end{center}
\caption{Structure Chart of an Algorithm to Compute Tractions from a
Solution to the Mild Slope Wave Equation.} \label{171}
\end{figure}
${}^1$ see Figure \ref{88} on page \pageref{88} for detail

${}^2$ see Figure \ref{89} on page \pageref{89} for detail

\section{Conclusions} 

The derivatives of a finite element solution can be successfully
computed on each element using 
\begin{eqnarray}  
\left. \left[ \frac{\partial f^h}{\partial x_1}, \ \frac{\partial
f^h}{\partial x_2}, \ \frac{\partial^2 f^h}{\partial x_1^2},
\ \frac{\partial^2 f^h}{\partial x_2^2}, \ \frac{\partial^2
f^h}{\partial x_1 \partial x_2} \right] \right|_{{\bx}({\bxi})} &=&
\sum^{\scriptsize \mbox{\it nNode}}_{i=1} c_i^{(e)} \left[ {\bQ}({\bxi})
\right]^{-1} {\bd}^i({\bxi})
\end{eqnarray}
where ${\bQ}({\bxi})$ and ${\bd}^i({\bxi})$ are defined in equation
(\ref{14}) on page \pageref{14}, $c_i^{(e)}$ is the discrete solution
on each element and $\mbox{\it nNode}$ is the number of nodes on each
element.

A bi--quadratic least squares fit (used to model the sea bed locally in
the vicinity of a node) can be calculated according to
\[
h(x,y) = [1, \ y, \ x, \ xy, \ y^2, \ x^2] \cdot {\bP}^{-1}{\bO} {\bz} 
\]
where ${\bP}$, ${\bO}$ are the matrices defined on page \pageref{141}
and ${\bz}$ is the vector of known depths used (sample points) in the
vicinity of the node in question. The various derivatives of this sea bed can be calculated using
\begin{eqnarray*}
\frac{\partial h}{\partial x} &=& [0, \ 0, \ 1, \ y, \ 0, \ 2x] \cdot
{\bf P}^{-1}{\bO} {\bz} \\
\frac{\partial h}{\partial y} &=& [0, \ 1, \ 0, \ x, \ 2y, \ 0] \cdot
{\bP}^{-1}{\bO} {\bz} \\
\frac{\partial^2 h}{\partial x^2} &=& [0, \ 0, \ 0, \ 0, \ 0,
\ 2] \cdot {\bP}^{-1}{\bf O}{\bz}  \\ 
\frac{\partial^2 h}{\partial y^2} &=& [0, \ 0, \ 0, \ 0, \ 2, \ 0]
\cdot {\bP}^{-1}{\bO} {\bz} \\
\frac{\partial^2 h}{\partial x \partial y} &=& [0, \ 0, \ 0, \ 1, \ 0,
\ 0] \cdot {\bP}^{-1}{\bO} {\bz}. 
\end{eqnarray*}
The components of the normal are then
\begin{eqnarray*} 
N_1(x,y) &=& - \frac{\partial h}{\partial x} \ = \ - [0, \ 0, \ 1, \ y,
\ 0, \ 2x] \cdot {\bP}^{-1}{\bO} {\bz}, \\ 
N_2(x,y) &=& - \frac{\partial h}{\partial y} \ = \ - [0, \ 1, \ 0, \ x,
\ 2y, \ 0] \cdot {\bP} ^{-1}{\bO}{\bz}, \\
N_3(x,y) &=& \frac{\partial h}{\partial h} \ = \ 1
\end{eqnarray*}
and the unit normal is 
\[
{\bn} (x,y) = \frac{{\bN} (x,y)}{\mid \mid {\bN} (x,y) \mid \mid_2}.
\]

A bi--quadratic least squares fit would appear to be a superior  method to
model the sea bed locally in the vicinity of a node when compared to the
more conventional approach which involves gridding and the use of cubic
splines.

The formula to compute the traction on the sea bed is given on page \pageref{175} (in terms of the derivatives of a finite element solution to
the mild slope wave equation and a least squares fitted bi--quadratic
model of the sea bed in the vicinity of each node). Lift, drag and Stokes' formula may all be calculated from elements incidental to it. 


\section{Appendix I}

\subsubsection*{The 8--Noded Quadrilateral Basis} \label{9} 

The basis used in conjunction with the 8--noded quadrilateral element is:
\begin{eqnarray*}
\phi_1({\bxi}) &=& \frac{1}{4} (\xi_1^2 - \xi_1)
(\xi_2^2 - \xi_2) \\ 
\phi_2({\bxi}) &=& \frac{1}{4} (\xi_1^2 + \xi_1)
(\xi_2^2 - \xi_2) \\
\phi_3({\bxi}) &=& \frac{1}{4} (\xi_1^2 + \xi_1)
(\xi_2^2 + \xi_2) \\
\phi_4({\bxi}) &=& \frac{1}{4} (\xi_1^2 - \xi_1)
(\xi_2^2 + \xi_2) \\
\phi_5({\bxi}) &=& - \frac{1}{2}
(\xi_1^2 - 1) (\xi_2^2 - \xi_2) \\ 
\phi_6({\bxi}) &=& - \frac{1}{2} (\xi_1^2 + \xi_1)
(\xi_2^2 - 1) \\
\phi_7({\bxi}) &=&  - \frac{1}{2} (\xi_1^2 - 1)
(\xi_2^2 - \xi_2) \\
\phi_8({\bxi}) &=& - \frac{1}{2} (\xi_1^2 - \xi_1)
(\xi_2^2 - 1)  
\end{eqnarray*}

\subsubsection*{The Derivatives of the 8--Noded Quadrilateral Basis}
\label{17}

The first derivatives of the 8--noded quadrilateral basis with respect to the first variable are:
\begin{eqnarray*}
\mbox{\it shape}(1,2) &\equiv& \frac{\partial \phi_1}{\partial \xi_1}
\ = \ \frac{1}{4}(\xi_2 + 2\xi_1 - 2\xi_1\xi_2 - \xi_2^2) \\ \mbox{\it
shape}(2,2) &\equiv& \frac{\partial \phi_2}{\partial \xi_1} \ = \ -
\xi_1 + \xi_1\xi_2 \\ \mbox{\it shape}(3,2) &\equiv& \frac{\partial
\phi_3}{\partial \xi_1} \ = \ \frac{1}{4}( - \xi_2 + 2\xi_1 -
2\xi_1\xi_2 + \xi_2^2) \\ \mbox{\it shape}(4,2) &\equiv& \frac{\partial
\phi_4}{\partial \xi_1} \ = \  \frac{1}{2}(1 - \xi_2^2) \\ \mbox{\it
shape}(5,2) &\equiv& \frac{\partial \phi_5}{\partial \xi_1} \ =
\ \frac{1}{4}(\xi_2 + 2\xi_1 + 2\xi_1\xi_2 + \xi_2^2) \\ \mbox{\it
shape}(6,2) &\equiv& \frac{\partial \phi_6}{\partial \xi_1} \ = \ -
\xi_1 - \xi_1\xi_2 \\ \mbox{\it shape}(7,2) &\equiv& \frac{\partial
\phi_7}{\partial \xi_1} \ = \ \frac{1}{4}( - \xi_2 + 2\xi_1 +
2\xi_1\xi_2 - \xi_2^2) \\ \mbox{\it shape}(8,2) &\equiv& \frac{\partial
\phi_8}{\partial \xi_1} \ = \  \frac{1}{2}(\xi_2^2 - 1).
\end{eqnarray*}
The first derivatives of the 8--noded quadrilateral basis with respect to
the second variable are:
\begin{eqnarray*}
\mbox{\it shape}(1,3) &\equiv& \frac{\partial \phi_1}{\partial \xi_2}
\ = \ \frac{1}{4}(\xi_1 + 2\xi_2 - 2\xi_1\xi_2 - \xi_1^2) \\ \mbox{\it
shape}(2,3) &\equiv& \frac{\partial \phi_2}{\partial \xi_2} \ =
\ \frac{1}{2}(\xi_1^2 - 1) \\ \mbox{\it shape}(3,3) &\equiv&
\frac{\partial \phi_3}{\partial \xi_2} \ = \ \frac{1}{4}( - \xi_1 +
2\xi_2 + 2\xi_1\xi_2 - \xi_1^2) \\ \mbox{\it shape}(4,3) &\equiv&
\frac{\partial \phi_4}{\partial \xi_2} \ = \ - (\xi_2 - \xi_1\xi_2) \\
\mbox{\it shape}(5,3) &\equiv& \frac{\partial \phi_5}{\partial \xi_2}
\ = \ \frac{1}{4}(\xi_1 + 2\xi_2 + 2\xi_1\xi_2 + \xi_1^2) \\ \mbox{\it
shape}(6,3) &\equiv& \frac{\partial \phi_6}{\partial \xi_2} \ =
\ \frac{1}{2}(1 - \xi_1^2) \\ \mbox{\it shape}(7,3) &\equiv&
\frac{\partial \phi_7}{\partial \xi_2} \ = \ \frac{1}{4}( - \xi_1 +
2\xi_2 - 2\xi_1\xi_2 + \xi_1^2) \\ \mbox{\it shape}(8,3) &\equiv&
\frac{\partial \phi_8}{\partial \xi_2} \ = \ (\xi_1\xi_2 - \xi_2).
\end{eqnarray*}
The second derivatives of the 8--noded quadrilateral basis with respect to
the first variable are:
\begin{eqnarray*}
\mbox{\it shape}(1,4) &\equiv& \frac{\partial^2 \phi_1}{\partial
\xi_1^2} \ = \ \frac{1}{2}(1 - \xi_2) \\ \mbox{\it shape}(2,4) &\equiv&
\frac{\partial^2 \phi_2}{\partial \xi_1^2} \ = \ \xi_2 - 1 \\ \mbox{\it
shape}(3,4) &\equiv& \frac{\partial^2 \phi_3}{\partial \xi_1^2} \ =
\ \frac{1}{2}(1 - \xi_2) \\ \mbox{\it shape}(4,4) &\equiv&
\frac{\partial^2 \phi_4}{\partial \xi_1^2} \ = \  0 \\ \mbox{\it
shape}(5,4) &\equiv& \frac{\partial^2 \phi_5}{\partial \xi_1^2} \ =
\ \frac{1}{2}(1 + \xi_2) \\ \mbox{\it shape}(6,4) &\equiv&
\frac{\partial^2 \phi_6}{\partial \xi_1^2} \ = \ - (1 + \xi_2) \\
\mbox{\it shape}(7,4) &\equiv& \frac{\partial^2 \phi_7}{\partial
\xi_1^2} \ = \ \frac{1}{2}(1 + \xi_2)\\ \mbox{\it shape}(8,4) &\equiv&
\frac{\partial^2 \phi_8}{\partial \xi_1^2} \ = \ 0.  \end{eqnarray*}
The second derivatives of the 8--noded quadrilateral basis with respect to
the second variable are:  
\begin{eqnarray*} \mbox{\it shape}(1,5) &\equiv& \frac{\partial^2
\phi_1}{\partial \xi_2^2} \ = \ \frac{1}{2}(1 - \xi_1) \\ \mbox{\it
shape}(2,5) &\equiv& \frac{\partial^2 \phi_2}{\partial \xi_2^2} \ = \ 0
\\ \mbox{\it shape}(3,5) &\equiv& \frac{\partial^2 \phi_3}{\partial
\xi_2^2} \ = \ \frac{1}{2}(1 + \xi_1) \\ \mbox{\it shape}(4,5) &\equiv&
\frac{\partial^2 \phi_4}{\partial \xi_2^2} \ = \ - (1 + \xi_1) \\
\mbox{\it shape}(5,5) &\equiv& \frac{\partial^2 \phi_5}{\partial
\xi_2^2} \ = \ \frac{1}{2}(1 + \xi_1) \\ \mbox{\it shape}(6,5) &\equiv&
\frac{\partial^2 \phi_6}{\partial \xi_2^2} \ = \ 0 \\ \mbox{\it
shape}(7,5) &\equiv& \frac{\partial^2 \phi_7}{\partial \xi_2^2} \ =
\ \frac{1}{2}(1 - \xi_1) \\ \mbox{\it shape}(8,5) &\equiv&
\frac{\partial^2 \phi_8}{\partial \xi_2^2} \ = \ \xi_1 - 1.
\end{eqnarray*}
The cross derivatives of the 8--noded quadrilateral basis are:
\begin{eqnarray*}
\mbox{\it shape}(1,6) &\equiv& \frac{\partial^2 \phi_1}{\partial \xi_1
\partial \xi_2} \ = \ \frac{1}{4}(1 - 2\xi_1 - 2\xi_2) \\ \mbox{\it
shape}(2,6) &\equiv& \frac{\partial^2 \phi_2}{\partial \xi_1 \partial
\xi_2} \ = \ \xi_1 \\ \mbox{\it shape}(3,6) &\equiv& \frac{\partial^2
\phi_3}{\partial \xi_1 \partial \xi_2} \ = \ \frac{1}{4}(2\xi_2 -
2\xi_1 - 1) \\ \mbox{\it shape}(4,6) &\equiv& \frac{\partial^2
\phi_4}{\partial \xi_1 \partial \xi_2} \ = \ - \xi_2 \\ \mbox{\it
shape}(5,6) &\equiv& \frac{\partial^2 \phi_5}{\partial \xi_1 \partial
\xi_2} \ = \ \frac{1}{4}(1 + 2\xi_1 + 2\xi_2) \\
 \mbox{\it shape}(6,6) &\equiv& \frac{\partial^2 \phi_6}{\partial \xi_1
 \partial \xi_2} \ = \ - \xi_1 \\ \mbox{\it shape}(7,6) &\equiv&
\frac{\partial^2 \phi_7}{\partial \xi_1 \partial \xi_2} \ =
\ \frac{1}{4}(2\xi_1 - 2\xi_2 -1) \\ \mbox{\it shape}(8,6) &\equiv&
\frac{\partial^2 \phi_8}{\partial \xi_1 \partial \xi_2} \ = \ \xi_2.
\end{eqnarray*}

\subsubsection*{The Various Derivatives of $Z(x_3,h)$} \label{18}

\[
\mbox{Since} \hspace{10mm} Z(x_3,h) = \frac{\cosh(\kappa(h
+ x_3))}{\cosh(\kappa h)} \hspace{10mm} \mbox{and}
\hspace{10mm} \frac{1}{\kappa} = {\mathop{\rm Tanh}}(\kappa h),
\]
\begin{eqnarray*}
\frac{\partial Z}{\partial x_1} &=& \frac{\kappa}{\cosh(\kappa h)} \left( \sinh(\kappa(h + x_3)) -
\frac{1}{\kappa} \cosh(\kappa(h + x_3)) \right)
\frac{\partial h}{\partial x_1} \\
& & + \ \frac{h}{\cosh(\kappa h)} \left( \sinh(\kappa(h + x_3)) - \frac{1}{\kappa} \cosh(\kappa(h +
x_3)) \right) \frac{\partial \kappa}{\partial x_1}, \\
& & \\
\frac{\partial Z}{\partial x_2} &=& \frac{\kappa}{\cosh(\kappa h)} \left( \sinh(\kappa(h + x_3)) -
\frac{1}{\kappa} \cosh(\kappa(h + x_3)) \right)
\frac{\partial h}{\partial x_2} \\
& & + \ \frac{h}{\cosh(\kappa h)} \left( \sinh(\kappa(h + x_3)) - \frac{1}{\kappa} \cosh(\kappa(h +
x_3)) \right) \frac{\partial \kappa}{\partial x_2}, \\
& & \\
\frac{\partial Z}{\partial x_3} &=& \kappa \frac{\sinh(\kappa(h + x_3))}{\cosh(\kappa h)}, \\
& & \\
\frac{\partial^2 Z}{\partial x_1^2} &=& \frac{\kappa}{\cosh(\kappa h)}
\left[ \left( \frac{1}{\kappa} + \kappa \right) \cosh(\kappa(h + x_3))
- 2 \sinh(\kappa(h + x_3)) \right] \left( \frac{\partial h}{\partial
x_1} \right)^2 \\
& & \\
& & + \ \frac{\kappa}{\cosh(\kappa h)} \left( \sinh(\kappa(h + x_3)) - \frac{1}{\kappa} \cosh(\kappa(h +
x_3)) \right) \frac{\partial^2 h}{\partial x_1^2} \\
& & \\
& & + \ \frac{h}{\cosh(\kappa h)} \left[ \left( 2 \kappa -
1 - \sinh(\kappa h) - \frac{1}{\kappa} - \frac{1}{h \kappa}
\right) \cosh(\kappa(h + x_3)) \right. \\
& & \\
& & \hspace{20mm} \left. + \left( 1 + \frac{2}{h} + \kappa \sinh(\kappa
h) - \kappa \right) \sinh(\kappa(h + x_3)) \right] \frac{\partial
h}{\partial x_1} \frac{\partial \kappa}{\partial x_1} \\
& & \\
& & + \ \frac{h}{\kappa \cosh(\kappa h)} \left[
\left(\frac{2}{\kappa} + h \kappa \right) \cosh(\kappa(h +
x_3)) - (1 + h) \sinh(\kappa(h + x_3)) \right] \left(
\frac{\partial \kappa}{\partial x_1} \right)^2 \\
& & \\
& & + \ \frac{h}{\cosh(\kappa h)} \left( \sinh(\kappa(h + x_3)) - \frac{1}{\kappa} \cosh(\kappa(h +
x_3)) \right) \frac{\partial^2 \kappa}{\partial x_1^2}, \\
& & \\
\frac{\partial^2 Z}{\partial x_2^2} &=& \frac{\kappa}{\cosh(\kappa h)}
\left[ \left(\frac{1}{\kappa} + \kappa \right) \cosh(\kappa(h + x_3)) -
2 \sinh(\kappa(h + x_3)) \right] \left( \frac{\partial h}{\partial x_2}
\right)^2 \\
& & \\
& & + \ \frac{\kappa}{\cosh(\kappa h)} \left( \sinh(\kappa(h + x_3)) - \frac{1}{\kappa} \cosh(\kappa(h +
x_3)) \right) \frac{\partial^2 h}{\partial x_2^2} \\
& & \\
& & + \ \frac{h}{\cosh(\kappa h)} \left[ \left( 2 \kappa -
1 - \sinh(\kappa h) - \frac{1}{\kappa} - \frac{1}{h \kappa}
\right) \cosh(\kappa(h + x_3)) \right. \\
& & \\
& & \hspace{20mm} \left. + \left( 1 + \frac{2}{h} + \kappa \sinh(\kappa
h) - \kappa \right) \sinh(\kappa(h + x_3)) \right] \frac{\partial
h}{\partial x_2} \frac{\partial \kappa}{\partial x_2} \\
& & \\
& & + \ \frac{h}{\kappa \cosh(\kappa h)} \left[
\left(\frac{2}{\kappa} + h \kappa \right) \cosh(\kappa(h +
x_3)) - (1 + h) \sinh(\kappa(h + x_3)) \right] \left(
\frac{\partial \kappa}{\partial x_2} \right)^2 \\
& & \\
& & + \ \frac{h}{\cosh(\kappa h)} \left( \sinh(\kappa(h + x_3)) - \frac{1}{\kappa} \cosh(\kappa(h +
x_3)) \right) \frac{\partial^2 \kappa}{\partial x_2^2}, \\
& & \\
\frac{\partial^2 Z}{\partial x_3^2} &=& \left( \kappa \right)^2
\frac{\cosh(\kappa(h + x_3))}{\cosh(\kappa h)},
\\
& & \\
\frac{\partial^2 Z}{\partial x_1 \partial x_2} &=&
\frac{\kappa}{\cosh(\kappa h)} \left[ \left(
\frac{1}{\kappa} + \kappa \right) \cosh(\kappa(h + x_3)) -
2 \sinh(\kappa(h + x_3)) \right] \frac{\partial h}{\partial
x_1}  \frac{\partial h}{\partial x_2} \\
& & \\
& & + \ \frac{\kappa}{\cosh(\kappa h)} \left( \sinh(\kappa(h + x_3)) - \frac{1}{\kappa} \cosh(\kappa(h +
x_3)) \right) \frac{\partial^2 h}{\partial x_1 \partial x_2} \\
& & \\
& & + \ \frac{h}{\cosh(\kappa h)} \left[ \left( \kappa -
\frac{1}{\kappa} - \frac{1}{h \kappa} \right) \cosh(\kappa(h + x_3))
\right. \\
& & \\
& & \hspace{20mm} \left. + \left( \frac{1}{h} - 2 \right)
\sinh(\kappa(h + x_3)) \right] \left( \frac{\partial h}{\partial x_1}
\frac{\partial \kappa}{\partial x_2} + \frac{\partial h}{\partial x_2}
\frac{\partial \kappa}{\partial x_1} \right) \\
& & \\
& & + \ \frac{h}{\kappa \cosh(\kappa h)} \left[
\left(\frac{2}{\kappa} + h \kappa \right) \cosh(\kappa(h +
x_3)) - (1 + h) \sinh(\kappa(h + x_3)) \right]
\frac{\partial \kappa}{\partial x_1} \frac{\partial \kappa}{\partial
x_2} \\
& & \\
& & + \ \frac{h}{\cosh(\kappa h)} \left( \sinh(\kappa(h + x_3)) - \frac{1}{\kappa} \cosh(\kappa(h +
x_3)) \right) \frac{\partial^2 \kappa}{\partial x_1 \partial x_2}, \\
& & \\
\frac{\partial^2 Z}{\partial x_1 \partial x_3} &=&
\frac{\kappa}{\cosh(\kappa h)} \left( \kappa \cosh(\kappa(h + x_3)) - \sinh(\kappa(h + x_3)) \right)
\frac{\partial h}{\partial x_1} \\
& & + \ \frac{h}{\cosh(\kappa h)} \left( \kappa \cosh(\kappa(h + x_3)) - \sinh(\kappa(h + x_3)) \right)
\frac{\partial \kappa}{\partial x_1} \hspace{10mm} \mbox{and} \\
& & \\
\frac{\partial^2 Z}{\partial x_2 \partial x_3} &=&
\frac{\kappa}{\cosh(\kappa h)} \left( \kappa \cosh(\kappa(h + x_3)) - \sinh(\kappa(h + x_3)) \right)
\frac{\partial h}{\partial x_2} \\
& & + \frac{h}{\cosh(\kappa h)} \left( \kappa \cosh(\kappa(h + x_3)) - \sinh(\kappa(h + x_3)) \right)
\frac{\partial \kappa}{\partial x_2}. 
\end{eqnarray*}

\nocite{dedecker:1}
\nocite{hockney:1}
\nocite{l:2}
\nocite{dbm:1}
\nocite{p:2}
\nocite{frank:1}

\bibliography{waves}

\end{document}